\documentclass[abstract=on,notitlepage,superscriptaddress,nofootinbib,aps,preprint,prd]{revtex4}
\usepackage{srcltx}
\usepackage{epsf}
\usepackage{amsthm}
\usepackage{amsfonts}
\usepackage{amsmath}
\usepackage{mathrsfs}
\usepackage{epsfig}
\usepackage{enumerate}
\usepackage{lipsum}
\usepackage{graphicx}
\usepackage{epsf}
\usepackage{subfig}
\usepackage{color}
\usepackage{caption}
\usepackage{subfig}
\usepackage[dvipsnames]{xcolor}
\usepackage{epstopdf}
\usepackage{float}
\usepackage{dcolumn}
\usepackage{hyperref}
\usepackage{amsthm}
\usepackage{graphicx, amsmath, amssymb, bm}

\newcommand{\be}{\begin{equation}}
\newcommand{\ee}{\end{equation}} 
\newcommand{\eei}{\end{equation}\indent\indent}
\newcommand{\bc}{\begin{center}}
\newcommand{\ec}{\end{center}}
\newcommand{\ber}{\begin{eqnarray*}}
\newcommand{\ear}{\end{eqnarray*}}
\newcommand{\ba}{\begin{array}}
\newcommand{\ea}{\end{array}}

\newcommand{\bea}{\begin{eqnarray}}
\newcommand{\eea}{\end{eqnarray}}

\newcommand{\ei}{\end{itemize}}

\newtheorem{theorem}{Theorem}

\begin{document}


 \title{Absence of curvature singularities in symmetric perfect fluid spacetimes in Einstein-Gauss-Bonnet Gravity}

\author{Aavishkar Madhunlall}\email[]{aavishkar.madhunlall@gmail.com}
\affiliation{Astrophysics Research Centre, School of Chemistry and Physics, University of KwaZulu--Natal, Private Bag X54001, Durban 4000, South Africa}
\author{Chevarra Hansraj}\email[]{hansrajc@sun.ac.za}
\affiliation{Applied Mathematics Division, Department of Mathematical Sciences, Stellenbosch University, Private Bag X1, Matieland 7602, South Africa}
\author{Rituparno Goswami}\email[]{goswami@ukzn.ac.za}
\affiliation{Astrophysics Research Centre, School of Mathematics, Statistics and Computer Science, University of KwaZulu--Natal, Private Bag X54001, Durban 4000, South Africa}
	\author{Sunil D. Maharaj}\email[]{maharaj@ukzn.ac.za}
\affiliation{Astrophysics Research Centre, School of Mathematics, Statistics and Computer Science, University of KwaZulu--Natal, Private Bag X54001, Durban 4000, South Africa}

\begin{abstract}
In this paper we study the higher dimensional homogeneous and isotropic perfect fluid spacetimes in Einstein-Gauss-Bonnet (EGB) gravity. We solve the modified field equations with higher order curvature terms to determine the evolution of the scale factor. We transparently show that this scale factor cannot become smaller than a finite minimum positive value which depends on the dimension and equation of state. This bound completely eliminates any curvature singularities in the spacetimes, where the scale factor must tend to zero. This is a unique property of EGB gravity which, despite being ghost-free and having quasi-linear field equations like general relativity, allows for the violation of singularity theorems. This phenomenon, thus, gives a natural way to dynamically construct regular black holes via higher dimensional continual gravitational collapse. 
 
\end{abstract}


\maketitle


\section{Introduction}
Higher dimensional cosmology, with initial roots in Kaluza-Klein \cite{Wesson1992} theory, has emerged as a significant area of interest in theoretical physics due to its recent implications on the late-time acceleration of spacetime \cite{Pahwa2011}, early spacetime \cite{Ranjit2012} and quantum gravity \cite{Kumar2024}. The Einstein-Gauss-Bonnet (EGB) gravity framework \cite{Lovelock1971, Boulware1985}, as a natural extension of general relativity (GR), provides an approach to explore the effects of higher order curvature corrections in spacetime dynamics. Unlike standard GR, the EGB theory incorporates a quadratic curvature term, which is topological in 4-dimensions (4D) but contributes dynamically in higher dimensions \cite{Maeda2006}. This makes EGB gravity particularly appealing for studying the early Universe, where such corrections may play a crucial role.

EGB cosmology has been studied via the modified Friedman equations in the context of the brane-world \cite{Deruelle2001, Charmousis2002, Maeda2003, Kofinas2005}, an alternative to dark energy \cite{Nojiri2005, Nojiri2007} and cosmic acceleration \cite{Deffayet2002}. Lately these types of studies comment on dynamics such as the expansion and inflation of spacetime \cite{Fomin2018, Gomez2022, Anjos2025} and even predict bouncing spacetimes (avoiding the Big Bang singularity) or modified black hole constructs. 

The consideration of perfect fluids in cosmology is motivated by their ability to model the matter-energy content of spacetime in a way that simplifies the founding equations. In the context of $N$-dimensional (ND) EGB cosmology, the perfect fluid model has been adopted in many studies already \cite{Charmousis2002, Maeda2006, Kofinas2005, Nojiri2005}.  Notably these fluids have no shear stresses, viscosity or heat conduction. Hence, it appears to be a good description of the observed spacetime on a large scale. Perfect fluids are generally used to describe radiation, matter, and even dark energy in cosmological models. 

Clearly the interplay between perfect fluids and higher dimensional EGB gravity is a promising framework for addressing key cosmological questions, such as the nature of singularities, the occurrence of bounces, and the evolution of spacetime under nonstandard conditions. This paper aims to extend the understanding of ND cosmology within the EGB framework, focusing on solutions that avoid singularities and lead to physically viable models. Specifically, we explore the implications of different equations of state for the matter content and discuss their consequences on the cosmological dynamics.

We comment briefly on the recent studies of homogeneous and isotropic spacetimes in EGB gravity. Singh \textit{et. al.} \cite{singh1,singh2} studied dust models in this context and obtained the implicit and explicit forms for the scale factor. Also solutions were found by imposing a linear equation of state for the standard matter. Our approach here is different and we utilise the conservation equation to write the scale factor and the matter variables, which enables us to consider general equations of state. 

The paper is organized as follows. In section \ref{sect2}, we provide an overview of ND EGB gravity. Then we introduce the founding cosmological Friedmann-Lema\^{i}tre-Robertson-Walker (FLRW) equations for modeling homogenous and isotropic spacetimes and explore its solutions in general relativity in section \ref{sect3}. We then extend these investigations to the EGB framework with solutions for $N>4$. After considering the energy conservation equation,  we use the enthalpy function to transform the field equations and find implicit solutions for cosmic time. We comment on black hole formation, singularities and gravitational collapse via a theorem in section \ref{sect5}. As applications of the theorem, we consider different forms of the perfect fluid equation of state in section \ref{sect6}. In particular we consider a general form of both linear and nonlinear equations of state, with Chaplygin gas falling under the latter category. Finally we perform analysis via plots of the scale factor as well as the gravitational constant and discuss the results in section \ref{sect7}.


\section{EGB gravity in $N$-dimensions} \label{sect2}
The Lovelock action extends the Einstein-Hilbert action by including higher order terms constructed from the Riemann curvature tensor and its contractions, which is written as
\begin{equation}
S = \int d^N x \sqrt{-g} \sum_{m=0}^{K} \alpha_m \mathcal{L}_m, \label{lovelock action}
\end{equation}
where $N$ is the spacetime dimension, $K = \lfloor (N-1)/2 \rfloor$ is the maximum order of nontrivial Lovelock terms, $\alpha_m$ are coupling constants and $\mathcal{L}_m$ represents the $m$-th order Gauss-Bonnet (GB) term.
The field equations are derived from \eqref{lovelock action} in the following form
\begin{equation}
\sum_{m=0}^K \alpha_m G_{ab}^{(m)} = \kappa T_{ab},
\end{equation}
where $G_{ab}^{(m)}$ terms are contributions from the $m$-th Lovelock term and $T_{ab}$ is the energy momentum tensor of the matter distribution. For each $m$, the term $G_{ab}^{(m)}$ is given by
\begin{equation}
G_{ab}^{(m)} = -\frac{1}{2} g_{ab} \mathcal{L}_m + m \left( R_{a}^{c_s d_1 e_1} \cdots R_{c_{m-1} d_{m-1} e_{m-1} b} - \frac{1}{2} g_{ab} \mathcal{L}_m \right).
\end{equation}
The equations of motion remain second order, avoiding the Ostrogradsky instability common in higher derivative theories \cite{padmanabhan2013} and suggesting further compatibility with classical GR \cite{Dadhich2016, Gannouji2019}.

EGB gravity corresponds to the $m=2$ case of Lovelock gravity in \eqref{lovelock action} with the second order GB term
\begin{equation} \label{LGB}
\mathcal{L}_{\text{GB}} = R_{abcd}\, R^{abcd} - 4R_{ab}R^{ab} + R^2.
\end{equation}
Its field equations are given by
\begin{equation}
G_{ab} + \Lambda g_{ab} - \frac{\alpha}{2} H_{ab} = \kappa T_{ab},
\end{equation}
where $H_{ab}$ is derived from the GB term
\begin{equation}
H_{ab} = g_{ab} \mathcal{L}_{\text{GB}} - 4 R R_{ab} + 8 R_{ac} R^c_b + 8 R^{cd} R_{acbd} - 4 R_a^{cde} R_{bcde},
\end{equation}
which vanishes when $N<5$, $\Lambda$ is the cosmological constant and Newton's gravitational constant in $N$-dimension is given as
\begin{equation}
\kappa_N = \frac{2(N-2)\pi^{(N-1)/2}}{(N-3)\Gamma\left(\frac{N-1}{2}\right)}\;.\label{kappa}
\end{equation}
The energy momentum tensor $T_{ab}$ for a perfect fluid is given by
\begin{equation}
T_{ab}=(\mu+p)u_au_b+pg_{ab}\,,
\end{equation}
where $u^a$ is the unit timelike vector tangent to the fluid flow lines. Also we impose all the physically realistic energy conditions on the matter fields and assume $\mu>0$ and $p>0$. 
 EGB gravity has been widely studied for its applications in black hole physics, cosmology, and gravitational wave astronomy \cite{fernandes2022}. In the limit where $\alpha \to 0$ in \eqref{lovelock action}, Einstein gravity in $N$-dimensions is regained causing the second order GB term \eqref{LGB} to have no contribution. We also take $\alpha>0$ to avoid any ghost instabilities in the theory.

\section{$N$-dimensional homogeneous and isotropic spacetimes} \label{sect3}
\subsection{FLRW field equations}
The FLRW metric serves as the foundation for modeling homogeneous and isotropic spacetimes. In $N$-dimensions, this metric generalizes to
\begin{equation}
ds^2 = -dt^2 + \frac{a^2(t)}{\left(1 + kr^2\right)^2}\left(dr^2 + r^2\left(d\Omega_{N-2}\right)^2\right),
\end{equation}
where
\begin{equation}
\left(d\Omega_{N-2}\right)^2 = \sum^{N-2}_{i=1} \left(\left[\prod^{i-1}_{j=1}\left(\sin^2 \theta_j\right)\right]\left(d\theta_i\right)^2\right),
\end{equation}
and  $k = 0, 1, -1$ are the spatial curvature constants. 

The scale factor, $a(t)$, represents the curvature and describes the angular part of the metric in $N$-dimensions. This generalization is crucial for understanding how different spatial curvatures impact the evolution of higher dimensional spacetimes. In the context of EGB gravity, the additional terms proportional to the coupling constant $\alpha$ reflect the influence of quadratic curvature corrections. These terms significantly alter the dynamics of the scale factor, particularly at high curvature scales. In observational cosmology, spatial curvature impacts the geometry of spacetime. For example, $k=0$ corresponds to a flat spacetime consistent with the $\Lambda$-Cold Dark Matter ($\Lambda$-CDM) model. Meanwhile, $k= \pm 1$ represents closed and opened spacetimes respectively, offering insights into different scenarios for cosmic evolution. For an extensive discourse on the different cases of spatial curvature in 4D perfect fluid cosmological spacetimes, the reader is referred to \cite{pramana}.
 
Now the $N$-dimensional field equations are given by
\begin{eqnarray} \label{nfe1}
\kappa_{N} \mu + \Lambda &=& \frac{(N-1)(N-2)}{2a^2}\left(\dot{a}^2+4k\right) \nonumber\\
& &+ \frac{\alpha(N-1)(N-2)(N-3)(N-4)}{2a^4}\left(\dot{a}^4+8\dot{a}^2k + 16k^2\right),
\end{eqnarray}
and
\begin{eqnarray} \label{nfe2}
-\kappa_{N} p + \Lambda &=& \frac{(N-2)}{a^2}\left[\frac{(N-3)}{2}\dot{a}^2 + a\ddot{a} + 2(N-3)k\right]  \\
&+ & \frac{2\alpha(N-2)(N-3)(N-4)}{a^3}\left[\ddot{a}\left(\dot{a}^2 + 4k\right)+\frac{(N-5)}{4a}\left(\dot{a}^4 + 8\dot{a}^2k + 16k^2\right)\right], \nonumber
\end{eqnarray}
where we regain the 4-dimensional general relativity field equations by substituting $N=4$. 
Following \cite{pramana}, we impose the equation of state 
\begin{equation} \label{eos}
S(\mu, p) = 0,
\end{equation}
and infer that 
\begin{equation}
\left(\frac{\partial S(\mu, p)}{\partial \mu}\right)^2 + \left(\frac{\partial S(\mu, p)}{\partial p}\right)^2 > 0.
\end{equation}
We reasonably assume that
\begin{equation}
\frac{\partial S(\mu, p)}{\partial p} \neq 0,
\end{equation}
and hence by the implicit function theorem \cite{textbook}, there exists a function $p(\mu)$ such that
\begin{equation} 
p=p(\mu),
\end{equation}
and
\begin{equation}
\frac{dp}{d\mu} = - \left[\frac{\partial S(\mu, p)}{\partial \mu} \Bigg/ \frac{\partial S(\mu, p)}{\partial p} \right].
\end{equation}
We further assume the following physically appropriate inequalities
\begin{equation}
a(t) > 0, \quad\quad \mu(t) > 0, \quad\quad p(t) > 0.
\end{equation}
The associated ND energy conservation equation from the field equations \eqref{nfe1} and \eqref{nfe2} is
\begin{equation} \label{ncons}
\dot{\mu} + (N-1)\frac{\dot{a}}{a}\left(\mu + p\right) = 0.
\end{equation}

\subsection{Solutions in general relativity 
($\alpha=0$)}

In GR, the dynamics of  ND homogeneous and isotropic spacetimes are governed by the Einstein field equations, which reduce to the FLRW equations for the metric specified above. The GR field equations are given by
\begin{eqnarray} \label{0fe1}
\kappa_{N} \mu + \Lambda &=& \frac{(N-1)(N-2)}{2a^2}\left(\dot{a}^2+4k\right), 
\end{eqnarray}
and
\begin{eqnarray} \label{0fe2}
-\kappa_{N} p + \Lambda &=& \frac{(N-2)}{a^2}\left[\frac{(N-3)}{2}\dot{a}^2 + a\ddot{a} + 2(N-3)k\right].
\end{eqnarray}
Taking a linear combination of the field equations $\left[\eqref{0fe1} - \left(N-1\right) \times \eqref{0fe2}\right]$ we obtain an expression for the cosmic acceleration 
\begin{eqnarray} \label{0raychau}
\ddot{a} &=& \frac{\left(-4 \left(N-2\right)^2 \left(\frac{\dot{a}^{2}}{4}+k\right) +2 a^{2} \left(-\kappa  p+\Lambda \right)\right) (N-1)+2 a^{2} \left(\kappa  \mu +\Lambda \right)}{2 a (N-1)(N-2)},
\end{eqnarray}
which highlights the influence of the $\kappa$-term in modifying the expansion dynamics. Substituting \eqref{0raychau} into field equation \eqref{0fe2} we get
\begin{equation}
\frac{4 \left(\frac{\dot{a}^{2}}{4}+k\right)(N-1)(N-2)-2 a^{2} \left(\kappa  \mu +\Lambda \right)}{2 a^{2} (N-1)} = 0,
\end{equation}
which when solving for $\dot{a}^2$ gives
\begin{equation} \label{ndota}
\dot{a}^2= \frac{2 a^{2}\left( \kappa  \mu +\Lambda\right)}{(N-1)(N-2)}-4k,
\end{equation}
Solving the field equations for the function $a(t)$ yields
\begin{equation}
\frac{dt}{da} = \left[ \frac{2 a^{2}\left( \kappa  \mu +\Lambda\right)}{(N-1)(N-2)}-4k\right]^{-\frac12}
\end{equation}
which leads to the integral
\begin{equation} \label{0integral}
t = \int  \left[ \frac{2 a^{2}\left( \kappa  \mu +\Lambda\right)}{(N-1)(N-2)}-4k\right]^{-\frac12} da + t_0,
\end{equation}
where $t_0$ is an integration constant and we regain the 4-dimensional general relativity results \cite{pramana} by substituting $N=4$.


\subsection{Solutions in EGB gravity ($\alpha\neq 0$)} \label{sect4}
The general ND field equations were previously given in \eqref{nfe1} and \eqref{nfe2}. In a similar manner as the previous subsection, taking a linear combination of the field equations $\left[\eqref{nfe1} - \left(N-1\right) \times \eqref{nfe2}\right]$ we obtain an expression for the cosmic acceleration in $N$-dimensions as
\begin{eqnarray} \label{nraychau}
\ddot{a} &=& \frac{1}{2 a(N-1)(N-2) \left(2 \alpha \left(\dot{a}^2+4 k\right) (N-3)(N-4)+a^{2}\right)} \times \nonumber\\
& & \left\{ -(N-1)\left[\left(\dot{a}^2+4 k\right) \left(\left(a^{2}+\alpha (N-4)\left((N-5)-1\right) \left(\dot{a}^{2}+4 k\right)\right) (N-3)+a^{2}\right) (N-2)\right] \right. \nonumber\\
&& \left. +2a^4\left[\kappa\left(\mu + (N-1)p\right) - (N-2) \Lambda\right]\right\}.
\end{eqnarray}
Substituting \eqref{nraychau} into field equation \eqref{nfe2} we get
\begin{equation}
\frac{-4(N-1) (N-2) \left(\frac{\dot{a}^{2}}{4}+k\right) \left(a^{2}-4 \alpha  \left(\frac{\dot{a}^{2}}{4}+k\right) (N-3) (N-4)\right)-2 a^{4} \left(\kappa  \mu +\Lambda \right)}{2 a^{4}(N-1)} = 0,
\end{equation}
which results in a quadratic equation in $\dot{a}^2$ and we can solve it to get
\begin{equation} \label{ndota}
\dot{a}^2=\frac{-8 \alpha  k \tilde{\mathcal{N}}_4 +a^{2} \left(-\tilde{\mathcal{N}}_2 \pm \sqrt{-8 \alpha  \tilde{\mathcal{N}}_4\left( \kappa  \mu + \Lambda\right) +\tilde{\mathcal{N}}_2^{2}}\right)}{2 \alpha  \tilde{\mathcal{N}}_4 },
\end{equation}
where $\tilde{\mathcal{N}}_2 = (N-1)(N-2)$ and $\tilde{\mathcal{N}}_4 = \tilde{\mathcal{N}}_2 (N-3)(N-4)$. Solving the field equations for the function $a(t)$ yields
\begin{equation}
\frac{dt}{da} = \left[\frac{-8 \alpha  k \tilde{\mathcal{N}}_4 +a^{2}\left( -\tilde{\mathcal{N}}_2 \pm \sqrt{-8  \alpha  \tilde{\mathcal{N}}_4\left( \kappa  \mu + \Lambda\right) +\tilde{\mathcal{N}}_2^{2}}\right)}{2 \alpha  \tilde{\mathcal{N}}_4 }\right]^{-\frac12},
\end{equation}
which in turn leads to the integral
\begin{equation} \label{nintegral}
t = \int \left[\frac{-8 \alpha  k \tilde{\mathcal{N}}_4 +a^{2} \left(-\tilde{\mathcal{N}}_2   \pm    \sqrt{-8  \alpha  \tilde{\mathcal{N}}_4\left( \kappa  \mu + \Lambda\right) + \tilde{\mathcal{N}}_2^{2}}\right)}{2 \alpha  \tilde{\mathcal{N}}_4 }\right]^{-\frac12} da + t_0,
\end{equation}
where $t_0$ is an integration constant and we are considering $N > 4$.

\section{Absence of Curvature Singularities} \label{sect5}
The energy conservation equation \eqref{ncons} encapsulates the evolution of energy density. By introducing the enthalpy function \cite{relcol} we establish a framework for connecting the thermodynamic properties of the fluid to its temporal evolution. This highlights the adiabatic nature of the fluid, consistent with the assumption of homogeneous and isotropic spacetimes. We define the enthalpy function \cite{pramana, relcol} as
\begin{equation} \label{M}
M(\mu) := \exp\left[\int \frac{d\mu}{\mu + p}\right] > 0,
\end{equation}
where $p = p(\mu)$. This implies
\begin{eqnarray}
\frac{dM(\mu)}{d\mu} = \frac{M(\mu)}{\mu + p} > 0,
\end{eqnarray}
since $M(\mu)>0, \quad \mu>0, \quad p > 0$. Now, since 
\begin{eqnarray}
\ln M(\mu) &=& \int \frac{d\mu}{\mu + p}\,,
\end{eqnarray}
we can use the ND energy conservation equation \eqref{ncons} to write the above as
\begin{eqnarray}
\frac{d}{dt}\left[\ln\left(a^{(N-1)} M(\mu)\right)\right] &=& 0, 
\end{eqnarray}
which implies
\begin{eqnarray}
 a^{(N-1)} M(\mu)&=& m_0,
\end{eqnarray}
where $m_0$ is a positive integration constant. By the inverse function theorem \cite{textbook}, the inverse function $M^{-1}$ exists such that 
\begin{eqnarray}
\label{dens}
\mu(t) &=& M^{-1}\left(m_0 a^{-(N-1)}\right).
\end{eqnarray} 
We substitute \eqref{dens} directly into the integrals in \eqref{0integral} and \eqref{nintegral} to obtain the following solutions (when $\alpha \neq 0$ and $N>4$)
\begin{equation}
\label{tN}
t = \int \left[\frac{-8 \alpha  k \tilde{\mathcal{N}}_4 +a^{2} \left(- \tilde{\mathcal{N}}_2   \pm    \sqrt{-8 \alpha  \tilde{\mathcal{N}}_4\left( \kappa  M^{-1}\left(m_0 a^{-(N-1)}\right) + \Lambda\right) + \tilde{\mathcal{N}}_2^{2}}\right)}{2 \alpha  \tilde{\mathcal{N}}_4 }\right]^{-\frac12} da + t_0,
\end{equation}
where $t_0$ is the integration constant and the right hand side is a function of the scale factor $a$.

By looking at the form of the integral \eqref{tN} analytically, we can infer the following in ND EGB perfect fluid collapse:
\begin{enumerate}
\item It is clear from the conservation equation that as the scale factor $a(t)$ monotonically decreases either in the future or in the past,  $M^{-1}\left(m_0 a^{-(N-1)}\right)$ will monotonically increase. Therefore, there must be an epoch (either in the future or in the past), when the negative term under the square root crosses the positive constant term, that is  
\begin{equation}
\mu \equiv M^{-1}\left(m_0 a^{-(N-1)}\right) = \frac{(N-2)(N-1)}{8\kappa\alpha (N-4)(N-3)} - \frac\Lambda\kappa.
\end{equation}
It is quite obvious that the evolution of the model cannot continue (in the future or in the past) beyond this epoch as the integrand becomes complex. Thus the scale factor attains the minimum positive value $(a_{\text{min}}$) at this epoch and it cannot become smaller than this.
\item The aforementioned bound is true for any equation of state $p = p(\mu)$.  
\item  This bound on the scale factor completely eliminates any curvature singularities (in the future or in the past), where the scale factor must tend to zero. Hence, Big Bang or Big Crunch models are not admissible in these types of cosmologies. Similarly any homogeneous gravitational collapse, for example the Oppenheimer-Snyder collapse, does not terminate at the singularity.
\item This phenomenon is distinctly different from the ND general relativistic case where $M^{-1}$ grows unbounded.
\end{enumerate}
This, then completes the proof of the following important theorem:
\begin{theorem} \label{thm1}
An $N$-dimensional homogeneous and isotropic perfect fluid spacetime in Einstein-Gauss Bonnet gravity contains no curvature singularities either in the future or in the past, without violating any energy conditions.
\end{theorem} 

In order to illustrate the effectiveness of the theorem we now consider specific cases of the equation of state. This will allow us to define the minimum scale factor value and comment on the cosmic behaviour of the models. 


\section{Special Cases} \label{sect6}
\subsection{Linear equation of state}

Linear equations of state, written in the general form $p=c_s\,^2\mu$, where the coefficient $c_s\,^2$ is the isentropic speed of sound, are particularly useful for modeling different phases of spacetime. For example in 4-dimensions, radiation-like matter and dust-like matter \cite{Padmanabhan2002} are described using the constants $\frac13$ and 0, respectively. These equations provide tractable solutions that bring out the interplay between matter and curvature in ND settings. In this paper, the derived solutions for $a(t)$ in flat, open, and closed spacetimes demonstrate how the scale factor evolves under different curvature scenarios. These results generalize the well known 4D solutions to higher dimensions, providing a broader perspective on cosmic evolution.

We consider the general form of a linear equation of state between the energy density and pressure given explicitly by
\begin{eqnarray}
p &=& c_{s}{}^{2} \mu. \label{pEqn}
\end{eqnarray}
Hence by the density relation in \eqref{dens} we can write
\begin{equation}
\label{dens2}
\mu = M^{-1}\left(m_0 a^{-(N-1)}\right) = \left(1 + c_s{}^{2}\right)^{-1}\left[\left(m_0 a^{-(N-1)}\right)^{1 + c_s{}^{2}}\right],
\end{equation}
and from \eqref{pEqn} we have
\begin{eqnarray}
\label{dens3}
\mu &=&  \left(1 + c_s{}^{2}\right)^{-1}\left(m_0 a^{-(N-1)}\right)^{1 + c_s{}^{2}}.
\end{eqnarray}
When $\alpha \neq0$, the dynamics are profoundly influenced by the higher order corrections in EGB gravity. Despite the resulting integral forms for $t(a)$ being implicit, they reveal critical features such as the existence of a minimum scale factor $a_{\text{min}}$, according to Theorem \ref{thm1}. This behavior excludes singularities like the Big Bang and Big Crunch, favoring a bouncing cosmology \cite{Nojiri2017}.

\begin{figure}[H]
    \centering
    \includegraphics[width=0.75\linewidth]{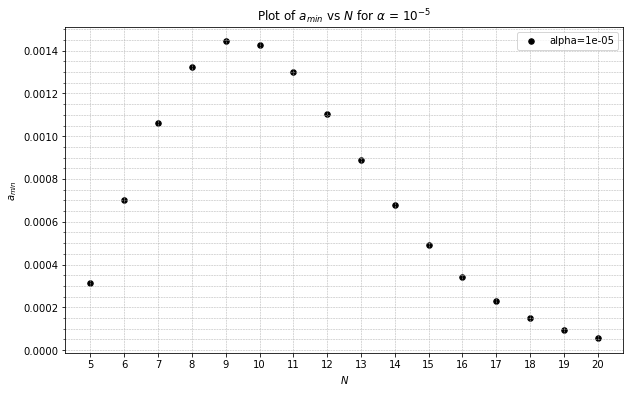}
    \caption{A plot of \eqref{a_0} showing $a_{\text{min}}$ vs $N$ with $N \in (5 ; 20)$}
    \label{fig1}
\end{figure}

\begin{figure}[H]
    \centering
    \includegraphics[width=0.75\linewidth]{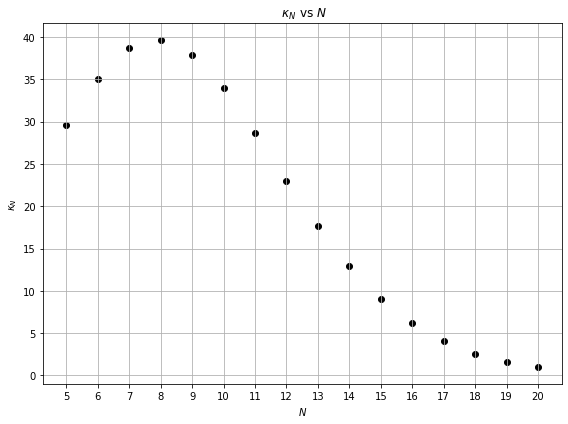}
    \caption{A plot of \eqref{kappa} showing $\kappa_N$ vs $N$.}
    \label{fig2}
\end{figure}

Repeating the same analytic procedure as before, considering \eqref{dens3}  the integral \eqref{tN} becomes
\begin{equation} \label{int1}
t = \int \left[\frac{-8 \alpha  k \tilde{\mathcal{N}}_4 +a^{2} \tilde{\mathcal{N}}_2   \pm    \sqrt{-8 a^{4} \alpha \kappa  \tilde{\mathcal{N}}_4 \left(1 + c_s{}^{2}\right)^{-1}\left(m_0 a^{-(N-1)}\right)^{1 + c_s{}^{2}} +a^{4} \tilde{\mathcal{N}}_2^{2}}}{2 \alpha  \tilde{\mathcal{N}}_4 }\right]^{-\frac12} da + t_0.
\end{equation}
We consider radiation-like matter. Performing the integration to obtain the $a(t)$ solution is complicated even if we consider the specific cases $k = 0, -1, +1$ due to the nature of the integrand. Hence we leave it in implicit form. However, we can still comment on the behaviour of $a(t)$. If we consider $a(t)$ in the integrand to be a decreasing function, there will exist some nonzero $a(t)$ beyond which the solution becomes imaginary. Another noticeable feature is that $a(t) \not\to 0$. Therefore, as stated in Theorem \ref{thm1}, Big Bang and Big Crunch models are not admissible hence there is no Oppenheimer-Snyder black hole. We can say that for any $a(t) < \left[\frac{8 \alpha \kappa  \tilde{\mathcal{N}}_4 \frac{N-1}{N}m_0{}^{N/N-1}}{ \tilde{\mathcal{N}}_2^{2}}\right]^{\frac1N} = a_0$, there is a minimum $a(t) = a_{min}$ beyond which the spacetime cannot continue. These models will give a bouncing solution for a finite $a(t)$.
By substituting in $\kappa_N$ \eqref{kappa} and making the assumption that $m_0=1$ for simplicity, we obtain
\begin{equation}
    a_{\text{min}}=\left(\frac{16(N-4)\alpha\pi^{(N-1)/2}}{N\left[\Gamma\left(\frac{N-1}{2}\right)\right]}\right)^{1/N}. \label{a_0}
\end{equation}
Dynamical analysis, numerical simulations and qualitative plots can further clarify the evolution of $a(t)$, particularly under varying values of $\alpha$ and curvature $k$ \cite{Fernandez2009, Bohmer2022}. These results emphasize the role of EGB gravity in stabilizing the cosmological evolution. Figure \ref{fig1} with $\alpha = 1\times10^{-5}$ depicts a sudden increase in the scale factor at $N=9$, a well known critical dimension \cite{paul}. This is because the energy density $\mu$ plays a crucial role in the governing of the expansion dynamics of the spacetime via the Friedmann equations. Particularly, a higher energy density around that dimension is indicated due to enhanced curvature effects. This suggests  $N=9$ is a topologically significant dimension where the GB contributions peak.
Figure \ref{fig2} presents a scatter plot of $\kappa_N$ \eqref{kappa} as a function of the number of spacetime dimensions $N$, for integer values ranging from 5 to 20. The peak at $N = 8$ after which there is a steady decline, underpins the effect of the gravitational constant $\kappa_N$ on dimensionality and the evolution of the model. Dimensionality plays a special role in amplifying the geometric contributions of higher-curvature terms, potentially influencing the structure of solutions, stability conditions, or compactification mechanisms in string-inspired models \cite{zwiebach1985}.

\subsection{Nonlinear equation of state: Chaplygin gas}
As an illustrative example of a nonlinear equation of state between the energy density and pressure, we consider a general Chaplygin gas model \cite{Kamenshchik2001, Bento2002}
\begin{equation}
\label{chapEOS}
S(\mu, p) = p - Z \mu + \frac{A}{\mu^w} = 0 \implies p = Z \mu - \frac{A}{\mu^w},
\end{equation}
where $Z, A$ and $w$ are prescribed constants.
From the definition of the function in \eqref{M} we will have,
\begin{equation}
\label{chapM}
M(\mu) = \exp\left[\int \frac{d\mu}{\mu\left(Z + 1\right)-A\mu^{-w}}\right] = \left[ \left(Z + 1\right)\mu^{w+1}-A\right]^{1/\nu},
\end{equation}
where $\nu = \left(Z + 1\right)\left(w + 1\right)$. We can then work out $\rho$, $p$ and finally obtain the integral 
\begin{equation}
t - t_0 = \int \left[\frac{-8 \alpha  k \tilde{\mathcal{N}}_4 +a^{2} \tilde{\mathcal{N}}_2   \pm    \sqrt{-8 a^{4} \alpha  \tilde{\mathcal{N}}_4\left( \kappa   \left\{\frac{1}{Z + 1}\left[A + \left(m_0 a^{-(N-1)}\right)^\nu\right]\right\}^{1/w+1} + \Lambda\right) +a^{4} \tilde{\mathcal{N}}_2^{2}}}{2 \alpha  \tilde{\mathcal{N}}_4 }\right]^{-\frac12} da,
\end{equation}
where the integral has to be left in implicit form. 

\section{Discussions} \label{sect7}
It is observed that in ND homogeneous and isotropic spacetimes governed by EGB gravity, curvature singularities do not form for physically reasonable equations of state, as stated in Theorem \ref{thm1}. This is plausible because EGB gravity can be considered to be general relativity with a correction term which we find is inducing an effective negative pressure geometrically and this is what is avoiding the singularity. The key results that emerged from our analysis are as follows.
\begin{enumerate}
\item We determine that the scale factor cannot become smaller than a finite minimum positive value which depends on the dimension and equation of state. By analyzing the linear equation of state integral \eqref{int1}, we can see that the minimum value of $a$ depends on $N$ and $\alpha$. Solving for $a$ within the inner square root gives us the expression \eqref{a_0} for \eqref{int1} to be real. By plotting a graph of $a$ vs $N$ for a fixed value of $\alpha$ in Figure \ref{fig1}, we observe that a peak occurs at $N=9$. This indicates that at this dimensionality, the minimum possible scale factor is maximized which suggests that the avoidance of singularities in EGB gravity depends on the number of spacetime dimensions $N$, with $N=9$ being the critical dimension where the minimum allowed scale factor reaches its largest value. 

\item The GB correction plays a role in shifting the dynamics of $a$, with its effects mostly pronounced at dimension $N=9$.  Therefore, this observed peak is an indication of an optimal dimension where the repulsive effects of the higher order curvature terms are maximized, preventing the formation of singularities most effectively. As a side note, we comment on the critical dimension $N=9$ arising in higher dimensional static stars in general relativity. With constant energy density, it was shown in \cite{paul}, that the mass to radius ratio reaches its maximum value at $N=9$. 

\item In the case of continual gravitational collapse, if $\alpha$ is extremely small, it is quite obvious that there would be a {\it bounce} of the collapsing ball within the trapped region. This is extremely interesting, as this violation of singularity theorem will give rise to regular black hole solutions, where the central singularity in a black hole would be prevented by higher dimensional Gauss-Bonnet geometry. 
 
\end{enumerate}

\begin{acknowledgements}
CH is supported by Stellenbosch University. AM, RG and SDM are supported by the National Research Foundation, South Africa and the University of KwaZulu-Natal.
\end{acknowledgements}

\end{document}